\documentclass[twocolumn,showpacs,preprintnumbers,amsmath,amssymb]{revtex4}
\usepackage{graphicx}% Include figure files
\usepackage{dcolumn}% Align table columns on decimal point
\usepackage{bm}% bold math

\begin{document}

\title{Three triton states in $^{9}$Li}
\author{K. Muta$^1$, T. Furumoto$^2$, T. Ichikawa$^2$, and N. Itagaki$^2$}

\affiliation{
$^1$Division of Physics and Astronomy, Kyoto University,
Kitashirakawa Oiwake-Cho, 606-8502 Kyoto, Japan}

\affiliation{
$^2$Yukawa Institute for Theoretical Physics, Kyoto University,
Kitashirakawa Oiwake-Cho, 606-8502 Kyoto, Japan}

\date{\today}% It is always \today, today,


\begin{abstract}
We focus on a characteristic non-$\alpha$ cluster structure
in light neutron-rich nuclei; three triton structure in $^9$Li.
This is an analogy to the case of three $\alpha$ state in $^{12}$C (Hoyle state).
The $\alpha$ clusters behave as bosons,
however tritons have Fermionic nature, 
and how the three cluster structure is different from $^{12}$C is an intriguing problem.
For this purpose, we introduce three triton wave functions.
In addition, $\alpha$+$t$+$n$+$n$ wave functions are prepared to describe 
other low-lying states of $^{9}$Li, and the coupling effect between them is taken into account.
The states with dominantly the three triton components appear below the three triton threshold energy,
where three triton correlation is important, however the root mean square radius is not 
enhanced contrary to the $\alpha$ gas states in $^{12}$C and $^{16}$O.

\end{abstract}


\maketitle

\section{Introduction}
Recently, gas-like states comprised of 
$\alpha$ clusters in atomic nuclei have attracted increased 
interest \cite{Tohsaki,Schuck,Funaki,Matsumura,Yamada, Kurokawa}.
One candidate is the second $0^+$ state of 
$^{12}$C$^*$ (3$\alpha$) at $E_x = 7.65$ MeV.
The state is well expressed by THSR 
(Tohsaki Horiuchi Schuck R\"{o}pke) wave function, 
where all of the $\alpha$-clusters
occupy the same $0s$-orbital with a spatially extended 
distribution \cite{Schuck}.
In addition, a 
candidate for the gas-like state of 4$\alpha$ clusters 
around the threshold energy in $^{16}$O has been studied
both by theoretical and experimental approaches \cite{Tohsaki,Funaki,Wakasa}.

Furthermore, not only $\alpha$ clusters, but also various other 
types of clusters 
 are expected to appear in the excited states of neutron-rich nuclei
 around the corresponding threshold energies.
One recent example is $\alpha$+$\alpha$+$t$ structure;
 the third $3/2^-$ state of $^{11}$B
 at $E_x = 8.56$ MeV has been
 discussed to have the $\alpha$+$\alpha$+$t$ structure 
 from both theoretical and experimental sides \cite{B11}.
Another example is $^{10}$Be; although
 all the low-lying states of $^{10}$Be have been found to be
 well described by the combination of three molecular-orbits
 for the two valence neutrons around two $\alpha$-clusters\cite{Ita-Be},
 the $\alpha$+$t$+$t$ cluster structure 
 is found to coexist
 with the $\alpha$+$\alpha$+$n$+$n$ structure
 around $E_x = 15$ MeV, close to the corresponding threshold\cite{Ita-Be2}.
The existence of $^7$Li+$t$ cluster 
 configuration of $^{10}$Be has been reported in the recent 
 experiment with $^8$Li$(d,t)^7$Li reaction \cite{Miyatake,Nelson},
 and since the $^7$Li nucleus is well-known to be described as
 $\alpha$+$t$ cluster structure, the resonance state
 is considered to have the $\alpha$+$t$+$t$ cluster configuration.
 This is one of the first example that 
 completely different cluster configurations 
 of triton-type ($\alpha+t+t$) and $\alpha$ type ($\alpha$+$\alpha$+$n$+$n$) 
 coexist in a single nucleus around the same energy region.
A well known other example is the appearance (mixing) of
 $^4$He+2$n$ and $t$+$t$ structures in $^6$He \cite{Csoto,Yamagata,Aoyama}.

In the present study, we focus on more drastic non-$\alpha$ cluster structure
in light neutron-rich nuclei; three triton structure in $^9$Li,
where all the clusters are tritons.
This is an analogy to the case of three $\alpha$ state in $^{12}$C (Hoyle state).
In the case of $\alpha$ cluster, it consists of four nucleons and 
the spin of the system is zero, thus it behaves as a boson.
On the other hand, a triton consists of three nucleons and has Fermionic nature, 
and how the three cluster structure is different from 
the Hoyle state in $^{12}$C is an intriguing problem.
Proposing an effective wave function for multi-triton (Fermion) systems,
just like THSR wave function in the $\alpha$ cluster case,
is the final goal.
In this paper, we focus on whether three triton states really appear or not in the
excited states of $^9$Li as the first step.
For this purpose, we introduce three triton wave functions.
Also $\alpha$+$t$+$n$+$n$ wave functions are introduced to describe 
other low-lying states of $^{9}$Li. It has been well known that $^7$Li
is described by a $\alpha$+$t$ model, thus $\alpha$+$t$+$n$+$n$
model is considered to work to large extent for the low-lying states of $^9$Li.
The validity of $\alpha$+$t$+$n$+$n$ model for $^9$Li has been shown in Ref.\cite{Arai},
and here, three triton states are coupled.

\section{Method}

We introduce two kind of the basis states; the basis states
with various $\alpha$+$t$+$n$+$n$ configurations,  $\{ \Psi(1)_i^{J^\pi KM} \}$,
and those with various $t$+$t$+$t$ configurations, $\{ \Psi(2)_j^{J^\pi KM} \}$.
%At first, we introduce $\alpha$+$t$+$n$+$n$ wave functions $\{ \Psi(1)_i^{J^\pi KM} \}$ 
%to describe the low-lying states of $^{9}$Li. 
It has been known that $^7$Li
is well described by $\alpha$+$t$, and $\alpha$+$t$+$n$+$n$
model is the first order approximation for the low-lying states of $^9$Li.
In addition, we introduce three triton wave functions $\{\Psi(2)_j^{J^\pi KM}\}$. 
The total wave function $\Phi^{J^\pi MK}$ is therefore
\begin{equation}
\Phi^{J^\pi KM} = \sum_i c_i \Psi(1)_i^{J^\pi KM} + \sum_j d_j \Psi(2)_j^{J^\pi KM}.
\end{equation}
The eigen states of Hamiltonian are obtained by diagonalizing 
the Hamiltonian matrix,
and the coefficients $\{c_i \}$ and $\{ d_j \}$
for the linear combination of each Slater determinant
are obtained.

The $i$-th basis state of $\{ \Psi(1)_i^{J^\pi KM} \}$ with 
the $\alpha$+$t$+$n$+$n$ configuration
has the following form,
\begin{eqnarray}
\Psi_i(1)^{J^\pi KM} = && P^\pi P^{JMK} \nonumber \\
&& [ {\cal A} \
\phi(\alpha, \vec r_1 \vec r_2 \vec r_3 \vec r_4, \vec R_1) \nonumber \\
&& \phi(t, \vec r_5 \vec r_6 \vec r_7, \vec R_2) \nonumber \\
&& \phi(n^{(1)}, \vec r_8, \vec R_3) \phi(n^{(2)}, \vec r_9, \vec R_4)]_i,
\end{eqnarray}
where $\cal A$ is the antisymmetrizer,
and 
$\phi(\alpha, \vec r_1 \vec r_2 \vec r_3 \vec r_4, \vec R_1)$, 
$\phi(t, \vec r_5 \vec r_6 \vec r_7, \vec R_2)$,
$\phi(n^{(1)}, \vec r_8, \vec R_3)$,  
$\phi(n^{(2)}, \vec r_9, \vec R_4)$ are
wave functions of $\alpha$, triton, the first valence neutron, and the second valence neutron,
respectively.
Here, $\{ \vec r_i \}$ represents spatial coordinates of nucleons, 
and each nucleon is described as locally shifted Gaussian centered at $\vec R$
with the size parameter of $\nu = 1/2b^2$, $b=$ 1.46 fm.
The $\alpha$ cluster consists of
four nucleons (spin-up proton, spin-down proton, spin-up neutron, and spin-down neutron),
which share a common Gaussian center parameter $\vec R_1$, although
the spin and isospin of each nucleon are not explicitly described in this formula
for simplicity. The triton consists of three nucleons 
(proton, spin-up neutron, and spin-down neutron) 
which are centered at $\vec R_2$.
The Gaussian center parameters of two valence neutrons are $\vec R_3$ and $\vec R_4$.
The $z$ components of the spins of the two valence neutrons are introduced to be
anti-parallel for simplicity.
The index $i$ in Eq. (2) specifies a set of values of Gaussian center parameters 
for $\vec R_1$, $\vec R_2$, $\vec R_3$, and $\vec R_4$.
The projection onto good parity and angular momentum 
(projection operators $P^\pi$ and $P^{JMK}$) is performed numerically. 
The number of mesh points for the Euler angle integral is 
%$24^3 = 13824$.
$16^3 = 4096$.

For the three triton basis sates,
the $j$-th basis state of $\{ \Psi(2)_j^{J^\pi KM} \}$ 
has the following form,
\begin{eqnarray}
\Psi_j(2)^{J^\pi KM} = && P^\pi P^{JMK} \nonumber \\
&& [ {\cal A} \
   \phi(t^{(1)}, \vec r_1 \vec r_2 \vec r_3, \vec R_1) \nonumber \\
&& \phi(t^{(2)}, \vec r_4 \vec r_5 \vec r_6, \vec R_2) \nonumber \\
&& \phi(t^{(3)}, \vec r_7 \vec r_8 \vec r_9, \vec R_3) ]_j,
\end{eqnarray}
where 
$\phi(t^{(1)}, \vec r_1 \vec r_2 \vec r_3, \vec R_1)$,
$\phi(t^{(2)}, \vec r_4 \vec r_5 \vec r_6, \vec R_2)$,  
$\phi(t^{(3)}, \vec r_7 \vec r_8 \vec r_9, \vec R_3)$ are
wave functions of three tritons.
In each triton cluster,
three nucleons  (proton, spin-up neutron, spin-down neutron)
share a common Gaussian center parameter $\{\vec R_1, \vec R_2$, or $\vec R_3 \}$.
The $z$ components of the spins of two of the protons are parallel,
and the remaining one is anti-parallel, for simplicity.

The Hamiltonian operator $(\hat{H})$ has the following form:
\begin{equation}
\displaystyle \hat{H}=\sum_{i= 1}^{A}\hat{t}_{i}-\hat{T}_{c. m.}+\sum_{i> j}^{A}\hat{v}_{ij}.
\end{equation}
 where the two-body interaction
 $(\hat{v}_{ij})$ includes the central, spin-orbit,
 and Coulomb parts.
As the $N-N$ interaction, for the central part, we use
 the Volkov No.2 effective potential \cite{Vol}:
\begin{eqnarray}
V( r)=&&( W-MP^{\sigma}P^{\tau}+ BP^{\sigma}-HP^{\tau})\nonumber \\&&
\times
( V_{1}\exp(-r^{2}/ c_{1}^{2})+ V_{2}\exp(-r^{2}/ c_{2}^{2})),
\end{eqnarray}
 where $c_{1}=$ 1.01 fm, $c_{2}= 1. 8$ fm, 
 $V_{1}=$ 61.14 MeV, $V_{2}=-60. 65$ MeV, 
 $W= 1-M$ and $M= 0. 60$.
The singlet-even channel
 of the original Volkov interaction without the
 Bartlet ($B$) and Heisenberg ($H$) parameters has been known to 
 be too strong, thus $B= H= 0.08$ is introduced to remove the bound
state of two neutrons.
For the spin-orbit term, we introduce the G3RS potential \cite{G3RS}:
\begin{equation}
V_{ls}= V_{0}( e^{-d_{1}r^{2}}-e^{-d_{2}r^{2}}) P(^{3}O){\vec{L}}\cdot{\vec{S}},
\end{equation}
 where $d_{1}= 5. 0$ fm$^{-2},\ d_{2}= 2. 778$ fm$^{-2}$,
 $V_{0}= 2000$ MeV,  and
 $P(^{3}O)$ is a projection operator onto a triplet odd state.
The operator $\vec{L}$ stands for the relative angular momentum
 and $\vec{S}$ is the spin ($\vec{S_{1}}+\vec{S_{2}}$).
All of the parameters of this interaction 
were determined from the $\alpha+ n$ and $\alpha+\alpha$
scattering phase shifts \cite{Okabe79}.

\section{Results}

We diagonalize the Hamiltonian comprised of the basis states
and obtain the eigen states and coefficients for the linear combination of
the basis states.
The energy convergence of $3/2^-$ states of $^{9}$Li is shown in Fig. 1.
From 1 to 150 on the horizontal axis are the wave functions
with various $\alpha$+$t$+$n$+$n$ configurations,  $\{ \Psi(1)_i^{J^\pi KM} \}$,
and from 151 to 300 are those with various $t$+$t$+$t$ configurations, 
$\{ \Psi(2)_j^{J^\pi KM} \}$. The ground state of $^9$Li converges 
at $-40.59$ MeV within the present framework.
This is 6.13 MeV below the $\alpha$+$t$+$n$+$n$ threshold 
(calculated at $-34.46$ MeV)
compared with the experimental value of 8.22 MeV.
We can see that the energies of some of the states drastically
decrease after superposing three triton configurations,
and eventually those states converge at $-30.66$ MeV as the third $3/2^-$ state,
$-29.25$ MeV as the 5th $3/2^-$ state,
and $-26.18$ MeV as the 7th $3/2^-$ state.
In this model, the three triton threshold is calculated 
at $-20.63$ MeV, corresponding to $E_x =$ 19.95 MeV,
compared with the experimental value of 19.55 MeV.
These states are considered to have predominantly 
the three triton configuration, which converges at the energy
much above the $\alpha$+$t$+$n$+$n$ threshold, 
however they are below the three triton threshold.
Therefore, the states are bound states with respect to the three triton 
threshold. 
The energy and root mean square (rms) radius are listed in Table I.
Although the third $3/2^-$ state has large excitation energy
of $E_x = 9.93$ MeV, it is much below the three triton threshold,
and the calculated rms radius of 2.39 fm is even smaller than
that of the ground state (2.43 fm).
Therefore, this nature should be called 
``three triton correlation" rather than ``three triton cluster"".
The 5th and 7th $3/2^-$ are closer to the threshold energy,
however the calculated rms radii are almost the same.

\begin{figure}
\includegraphics[width=\linewidth]{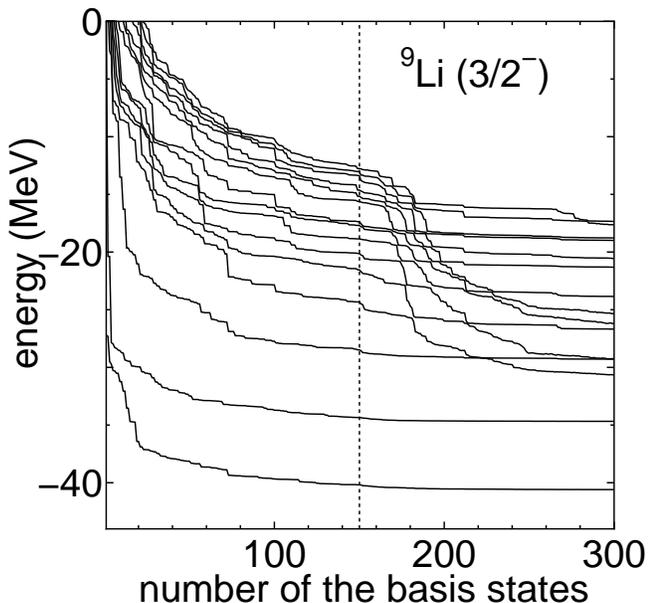}
\caption{
The energy convergence of $3/2^-$ states of $^{9}$Li.
From 1 to 150 on the horizontal axis are the wave functions
with various $\alpha$+$t$+$n$+$n$ configurations,  $\{ \Psi(1)_i^{J^\pi KM} \}$,
and from 151 to 300 are those with various $t$+$t$+$t$ configurations, 
$\{ \Psi(2)_j^{J^\pi KM} \}$.}
\end{figure}

\begin{figure}
\includegraphics[width=\linewidth]{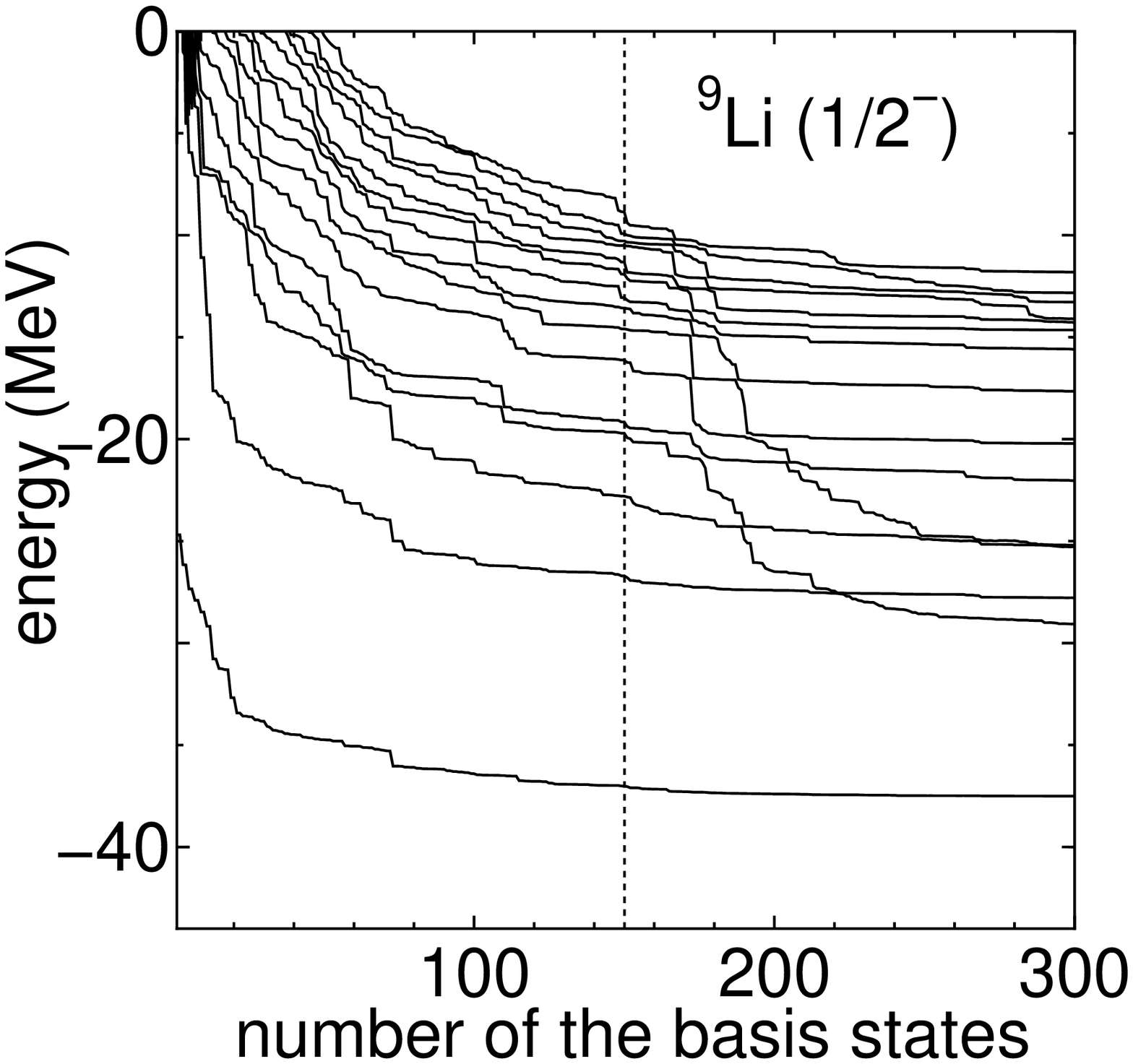}
\caption{
The energy convergence of $1/2^-$ states of $^{9}$Li.
The definition of the horizontal axis is the same as Fig. 1.
}
\end{figure}

Similar situation can be seen also for the $1/2^-$ states in Fig. 2.
The energies of two of the states which drastically
decrease after superposing three triton configurations
converge at $-29.06$ MeV as the second $1/2^-$ state
and $-25.27$ MeV as the 4th $1/2^-$ state.
The second  $1/2^-$ state is below the three triton threshold by 8.43 MeV, and 
the rms radius is also small (2.37 fm) similarly to the $3/2^-$ case.
The 4th $1/2^-$ state is closer to  the threshold, however
the calculated rms radius is 2.38 fm.
The lowest $1/2^-$ state is obtained at $-37.50$ MeV, 3.09 MeV above the
ground state (experimental value is $E_x$ = 2.691 MeV).

\begin{figure}
\includegraphics[width=\linewidth]{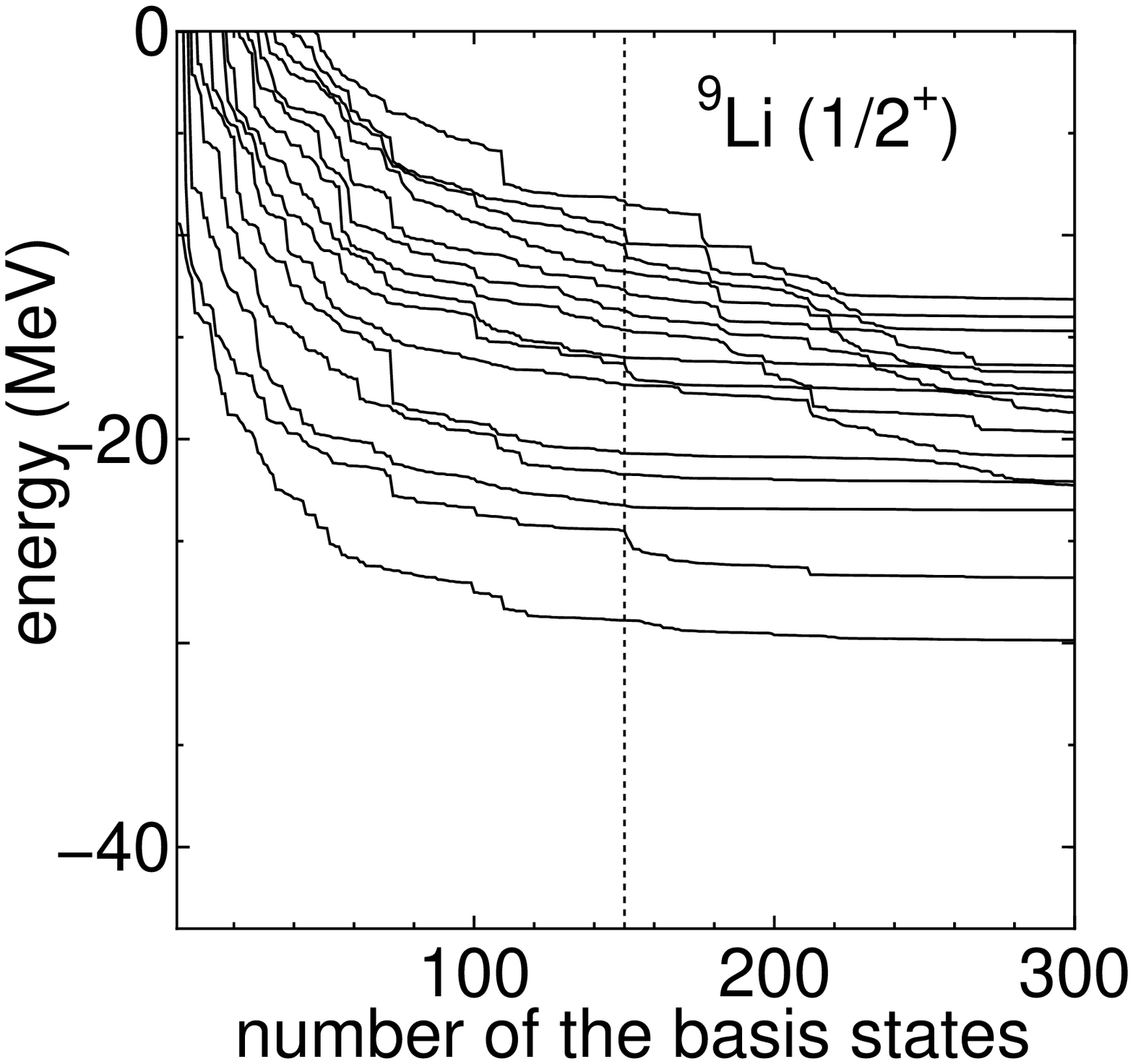}
\caption{
The energy convergence of $1/2^+$ states of $^{9}$Li.
The definition of the horizontal axis is the same as Fig. 1.
}
\end{figure}

The situation is slightly different in the $1/2^+$ case.
Figure 3 shows that the state largely affected by the three triton 
configurations converges as the 4th $1/2^+$ state at $-22.25$ MeV.
This is below the three triton threshold by 1.62 MeV, and the calculated radius 
is rather large, 2.62 fm, larger than that of the ground state by 0.19 fm.
In the $^9$Li case, positive parity states correspond to higher nordal states,
and this could be one of the reasons for the (slightly) large radius.
However, it is not enhanced compared with the neutron-halo structure in $^{11}$Li
or Hoyle state in $^{12}$C, and it is smaller than the yrast $1/2^+$ state (2.71 fm).
It is also shown that the second $1/2^+$ state is affected when
three triton states are coupled to the $\alpha$+$t$+$n$+$n$ basis states. 
Therefore, the mixing (or coupling) effect of two components, $\alpha$+$t$+$n$+$n$ 
and $t+t+t$ would be important in the second $1/2^+$ state.

\begin{figure}
\includegraphics[width=\linewidth]{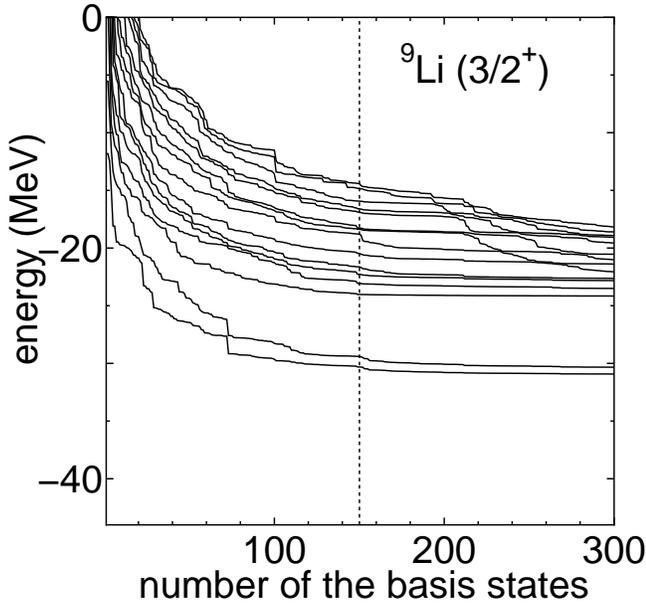}
\caption{
The energy convergence of $3/2^+$ states of $^{9}$Li.
The definition of the horizontal axis is the same as Fig. 1.
}
\end{figure}

\begin{table}
\begin{center}
\caption{
The calculated root mean square radius (fm)
of the $1/2^+$, $3/2^+$, $1/2^-$, and $3/2^-$ states.
The energies (MeV) are shown in the parentheses.
}
\begin{tabular}{||c|c|c|c|c|}
\hline
\hline
     & $1/2^+$  & $3/2^+$  &  $1/2^-$ & $3/2^-$ \\
\hline
1 & 2.71 ($-29.86$) & 2.72 ($-30.92$) & 2.53 ($-37.50$) & 2.43 ($-40.59$) \\
2 & 2.92 ($-26.79$) & 2.75 ($-30.33$) & 2.37 ($-29.06$) & 2.59 ($-34.68$) \\
3 & 2.62 ($-23.47$) & 2.66 ($-24.16$) & 2.78 ($-27.78$) & 2.39 ($-30.66$) \\
4 & 2.62 ($-22.25$) & 2.77 ($-23.50$) & 2.38 ($-25.27$) & 2.69 ($-29.29$) \\
5 & 2.71 ($-22.07$) & 2.85 ($-22.82$) & 2.84 ($-25.19$) & 2.38 ($-29.25$) \\
6 & 2.66 ($-20.83$) & 2.74 ($-22.64$) & 2.97 ($-22.04$) & 2.86 ($-26.67$) \\
7 & 3.11 ($-19.66$) & 2.60 ($-22.07$) & 2.88 ($-20.21$) & 2.34 ($-26.18$) \\
\hline
\end{tabular}
\end{center}
\end{table}

The result for the $3/2^+$ states is shown in Fig. 4, and
the 7th $3/2^+$ state is largely affected by the three triton configurations 
and converges at $-22.07$ MeV. This state is also close to the threshold,
however the calculated rms radius of 2.60 fm is not so strongly enhanced.

The component of three triton clusters for each state is listed in Table II.
Here the triton cluster component ($C^{3t}_i$) of the state $\phi_i$ is defined as
\begin{equation}
C^{3t}_i = \sum_k \langle \phi_i | \psi_k^{3t} \rangle \langle \psi_k^{3t} | \phi_i \rangle. 
\end{equation}
Here $\psi_k^{3t}$ is the $k$-th orthonormal state obtained by diagonalizing 
the Hamiltonian matrix only within the basis states of three triton clusters 
(from 151 to 300 on the horizontal axis of Figs. 1-4).
It can be confirmed that states discussed to be affected by
three triton basis states really have large three triton components;
the second $1/2^-$ and third $3/2^-$ states have the amount of
$\sim$1.00, and the 4th $1/2^-$ and 5th $3/2^-$ states 
have the amount of 0.96$\sim$0.99. 
The second $1/2^+$ has mixed structure;
$C^{3t}_i$ is around 0.76. 

\begin{table}
\begin{center}
\caption{
The components of three triton clusters  $C^{3t}_i$ defined in Eq. (7).
The values in the parentheses show the energies of the states.
}
\begin{tabular}{||c|c|c|c|c|}
\hline
\hline
     & $1/2^+$  & $3/2^+$  &  $1/2^-$ & $3/2^-$ \\
\hline
1 & 0.66 ($-29.86$) & 0.67 ($-30.92$) & 0.65 ($-37.50$) & 0.67 ($-40.59$) \\
2 & 0.76 ($-26.79$) & 0.75 ($-30.33$) & 1.00 ($-29.06$) & 0.77 ($-34.68$) \\
3 & 0.01 ($-23.47$) & 0.06 ($-24.16$) & 0.44 ($-27.78$) & 1.00 ($-30.66$) \\
4 & 0.88 ($-22.25$) & 0.30 ($-23.50$) & 0.99 ($-25.27$) & 0.40 ($-29.29$) \\
5 & 0.23 ($-22.07$) & 0.51 ($-22.82$) & 0.66 ($-25.19$) & 0.96 ($-29.25$) \\
6 & 0.08 ($-20.83$) & 0.23 ($-22.64$) & 0.69 ($-22.04$) & 0.68 ($-26.67$) \\
7 & 0.71 ($-19.66$) & 0.99 ($-22.07$) & 0.34 ($-20.21$) & 1.00 ($-26.18$) \\
\hline
\end{tabular}
\end{center}
\end{table}

In a similar way,
the component of $\alpha$+$t$+$n$+$n$ for each state is listed in Table III.
Here the $\alpha$+$t$+$n$+$n$ component ($C^{\alpha tnn}_i$) of the state $\phi_i$ is defined as
\begin{equation}
C^{\alpha tnn}_i = \sum_k \langle \phi_i | \psi_k^{\alpha tnn} \rangle \langle \psi_k^{\alpha tnn}
| \phi_i \rangle, 
\end{equation}
where $\psi_k^{\alpha tnn}$ is the $k$-th orthonormal state obtained by diagonalizing 
the Hamiltonian matrix only within the basis states of $\alpha$+$t$+$n$+$n$ configurations 
(from 1 to 150 on the horizontal axis of Figs. 1-4).
It can be confirmed that states affected by
the three triton states have small values;
the second $1/2^-$ and third $3/2^-$ states have the values less than 1 \%,
and the 4th $1/2^-$ and 5th $3/2^-$ states have 0.02 and 0.05, respectively.
The second and 4th $1/2^+$ states have mixed structure and the values are 0.86 and 0.15.
Since the wave functions of 
$\alpha$+$t$+$n$+$n$ and three triton configurations are non-orthogonal,
the sum of the values of Table II and Table III for each state
does not become unity.

\begin{table}
\begin{center}
\caption{
The components of $\alpha$+$t$+$n$+$n$ configurations $C^{\alpha tnn}_i$ defined in Eq. (8).
The values in the parentheses show the energies of the states.
}
\begin{tabular}{||c|c|c|c|c|}
\hline
\hline
     & $1/2^+$  & $3/2^+$  &  $1/2^-$ & $3/2^-$ \\
\hline
1 & 0.97 ($-29.86$) & 0.98 ($-30.92$) & 0.99 ($-37.50$) & 0.99 ($-40.59$) \\
2 & 0.86 ($-26.79$) & 0.96 ($-30.33$) & 0.00 ($-29.06$) & 0.99 ($-34.68$) \\
3 & 0.99 ($-23.47$) & 0.99 ($-24.16$) & 0.93 ($-27.78$) & 0.00 ($-30.66$) \\
4 & 0.15 ($-22.25$) & 0.94 ($-23.50$) & 0.02 ($-25.27$) & 0.89 ($-29.29$) \\
5 & 0.89 ($-22.07$) & 0.94 ($-22.82$) & 0.90 ($-25.19$) & 0.05 ($-29.25$) \\
6 & 0.93 ($-20.83$) & 0.95 ($-22.64$) & 0.81 ($-22.04$) & 0.90 ($-26.67$) \\
7 & 0.76 ($-19.66$) & 0.01 ($-22.07$) & 0.95 ($-20.21$) & 0.00 ($-26.18$) \\
\hline
\hline
\end{tabular}
\end{center}
\end{table}

\begin{table}
\begin{center}
\caption{
The calculated energy (MeV) and squared isoscalar E0  
transition strength from the ground state (fm$^4$)
for the $3/2^-$ states.
}
\begin{tabular}{||c|c|c|c|c|}
\hline
\hline
energy (MeV) & squared E0 (fm$^4$)  \\
\hline
 $-40.59$ & --- \\
 $-34.68$ & 0.03  \\
 $-30.66$ & 0.01 \\
 $-29.29$ & 40.51 \\
 $-29.25$ & 2.11 \\
 $-26.67$ & 21.68 \\
 $-26.18$ & 0.01 \\
\hline
\end{tabular}
\end{center}
\end{table}

Recently, it has been proposed that the strong enhancement 
of isoscalar monopole (E0) transitions can be a measure
of the existence of cluster structure \cite{Kawabata}.  
The isoscalar E0 operator has a form of $\sum_i r_i^2$, where
the sum over $i$ is for all the nucleons.
From the experimental side, the presence of cluster states in $^{13}$C
has been suggested by measuring the isoscalar E0 transitions
from the ground $1/2^-$ state induced by the $^{13}$C$(\alpha, \alpha')^{13}$C
reaction \cite{Sasamoto}. 
Also from the theoretical side, 
the relation between the monopole transition strength and the cluster
structure has been widely discussed \cite{YIO, Yoshida-2,Uegaki, HO, Yamada-1,Ichikawa}. 
The squared isoscalar E0 transition strength from the ground $3/2^-$ state
to low-lying $3/2^-$ states is listed in Table IV. Although the strength
to the 4th $3/2^-$ state shows an enhanced value of
$\sim$ 40 fm$^4$, the strengths to the third and 5th $3/2^-$ states,
which have large components of three tritons, are very suppressed.
This is considered to be due to the small rms radius of the third and 5th $3/2^-$ state,
and the other reason comes from the fact that
the wave functions of the ground and those $3/2^-$ states
have completely different structure.

\section{Conclusion}

We focused on the non-$\alpha$ cluster structure
in light neutron-rich nuclei; three triton structure in $^9$Li.
This is an analogy to the case of three $\alpha$ state in $^{12}$C (Hoyle state).
For this purpose, we introduced both three triton wave functions
and $\alpha$+$t$+$n$+$n$ wave functions for the low-lying states of $^{9}$Li,
and coupling effect has been taken into account. 

The calculated second $1/2^-$ and third $3/2^-$ states have 
dominantly the components of three triton clusters,
and the states are well below the threshold energy of three tritons.
The calculated rms radii are quite small, and the E0 transition strength from the ground
state to the third $3/2^-$ state is also small. 
Therefore, this nature should be called 
``three triton correlation" rather than ``three triton cluster"".
The situation is the same for other three triton states
close to the threshold.

On the other hand, the positive parity states correspond
to higher nordal states, and the states have slightly larger rms radii.
However, 
the mixing (or coupling) effect of two components, $\alpha$+$t$+$n$+$n$ 
and $t+t+t$ is important in these states.

Nevertheless, three triton correlation is important and states with
dominantly the three triton components appear in the low-lying
states of $^9$Li.

\begin{acknowledgments}
%The authors would like to thank C. Wheldon and P. Schuck for important suggestions. 
%Numerical computation in this work was carried out at the 
%Yukawa Institute Computer Facility using the new SR16000 system.
This work was done in the Yukawa International Project for Quark-Hadron Sciences (YIPQS).
\end{acknowledgments}

\end{document}